\documentstyle[11pt,epsf,newpasp,twoside]{article}
\def\edcomment#1{\iffalse\marginpar{\raggedright\sl#1\/}\else\relax\fi}
\reversemarginpar

\begin{document}
\title{The Galaxy Environment of Quasars in the ${\bf z\simeq1.3}$ Clowes-Campusano Large Quasar Group}
 \author{Haines C.P. \& Clowes R.G.}
\affil{Centre for Astrophysics, University of Central Lancashire, Preston, \\PR1 2HE, UK}
\author{Campusano L.E.}
\affil{Observatorio Astr\'{o}nomico Cerro Cal\'{a}n, Departamento de Astronom\'{i}a, Universidad de Chile, Casilla 36-D, Santiago, Chile}

\begin{abstract}
We report significant associated clustering in the field of a $z=1.226$ quasar from the Clowes-Campusano LQG in the form of both a factor $\sim11$ overdensity of $I-K>3.75$ galaxies, and red sequences of 15--18 galaxies at $I-K\simeq4.3,V-K\simeq6.9$ indicative of a population of massive ellipticals at the quasar redshift.
 The quasar is located between two groups of these galaxies, with further clustering extending over \mbox{2--3 Mpc}. A band of $V-I<1$ galaxies bisects the two groups of red sequence galaxies, and we suggest that the merging of these two groups has triggered both this band of star-formation and the quasar.
\end{abstract}

\section{Introduction}

Quasars have been used as efficient probes of high-redshift clustering because they are known to favour rich environments. Quasars may also trace large-scale structures at early epochs ($0.4\la z\la 2$) in the form of Large Quasar Groups (LQGs), which have comparable sizes to the largest structures at the present epoch. The largest of these is the Clowes-Campusano LQG of 18 quasars at $z\simeq1.3$, with a maximal extent of 200$h^{-1}\,$Mpc (Clowes \& Campusano 1991).

To examine the galaxy environment of quasars in this LQG, we have conducted an ultra-deep optical (to $V\sim27,I\sim26$) study of a $30'\times30'$ field containing 3 quasars from this LQG, with additional $K$ imaging obtained for subfields around 2 of them. Any associated passively-evolving galaxies can be identified from their extremely-red optical-IR colours ($I-K\simeq4,V-K\simeq7$).

\begin{figure}
\plotfiddle{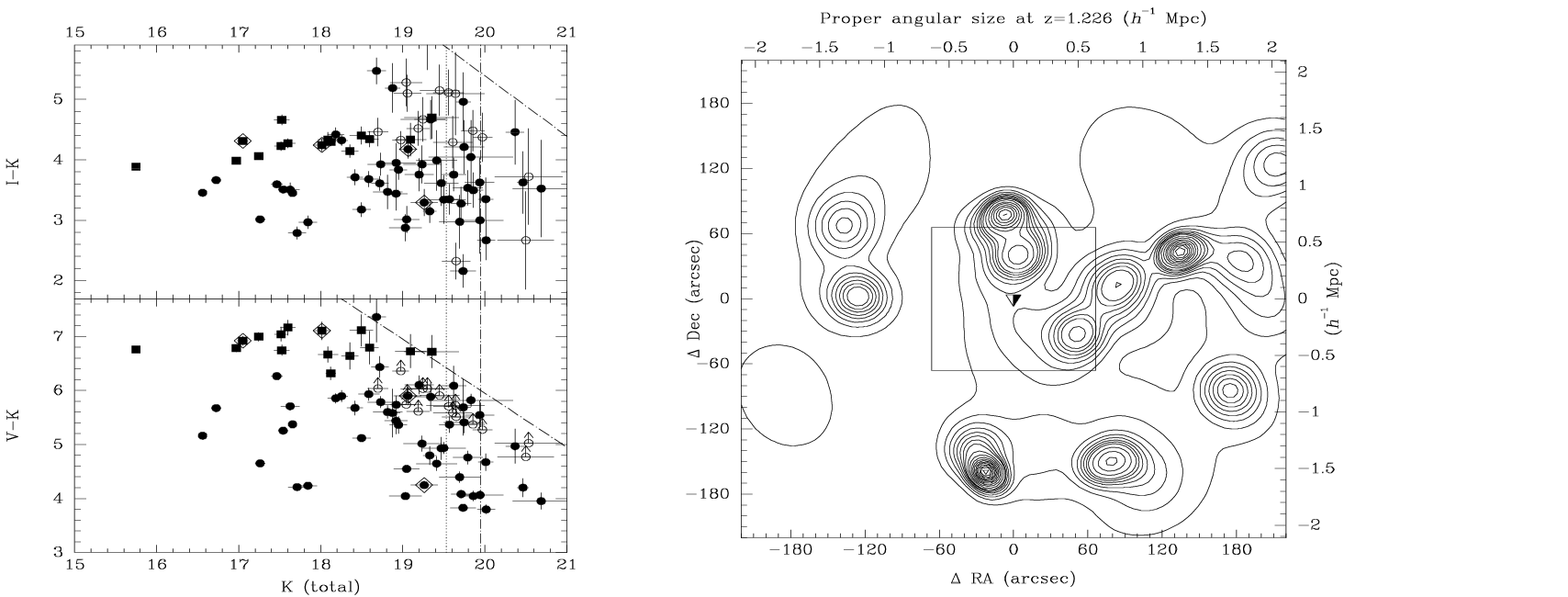}{2.1in}{0}{55}{55}{-310}{-140}
\caption{{\it Left:} C-M diagrams for the z=1.226 LQG quasar field.}
\caption{{\it Right:} Density distribution of $V-I>2.25,I<23$ galaxies around the z=1.226 quasar, showing associated large-scale structure.}
\end{figure}

\section{Galaxy Clustering Around a z=1.226 LQG Quasar}

We find a 3.5$\sigma$ excess of $K<20$ galaxies in the $2.25'\times2.25'$ \mbox{($1.3\times1.3h^{-2}\,{\rm Mpc}^{2}$)} field centred on the z=1.226 LQG quasar 104420.8+055739 (Haines et al. 2000). This excess is due entirely to a factor $\sim$11 overdensity of $I-K>3.75$ galaxies, which must have $z\ga0.8$ to explain their colour. We observe clear red sequences at \mbox{$V-K\simeq6.9,$} and $I-K\simeq4.3$ (Figure 1), comparable to red sequences observed in other $z\simeq1.2$ clusters, and indicating a population of 15--18 passively-evolving massive ellipticals at the quasar redshift.
 We have produced a density map (Figure 2) of galaxies with the same optical colours and magnitudes as the red sequence galaxies \mbox{($V-I>2.25,I<23$)} to estimate the full extent of any associated clustering beyond the $K$ image (shown as box). There is clearly significant substructure, with the quasar located between two groups of red sequence galaxies and further clustering can be seen extending over \mbox{2--3$h^{-1}\,$Mpc.} The whole structure is suggestive of being a cluster in the early stages of formation through the coalescence of subclusters.
The quasar is located within a concentration of blue ($V-I<1$) galaxies, with $\ga50\%$ clearly at $z\ga1$, having $I-K\ga3.5$, indicating a region of enhanced star-formation, similar to those found around several $z\simeq1.1$ LQG quasars (Hutchings, Crampton \& Johnson 1995). The concentration appears extended to form a {\em band} which bisects the two groups of red sequence galaxies. We propose that this band of recent star-formation, and the quasar itself, have been triggered by the merging of the two groups, as the galaxies interact with the colliding ICMs.

The two other LQG quasars are in poor environments, although a background $z=1.426$ quasar lies on a cluster periphery. Quasars at these redshifts ($1\la z\la1.5$) are found either on the peripheries of clusters (see also Sanch\'{e}z \& Gonz\'{a}lez-Serrano 1999) or in relatively undistinguished environments.

\end{document}